\documentclass[%
 reprint,
superscriptaddress,
 amsmath,amssymb,
 aps,prl,
]{revtex4-2}

\usepackage{graphicx}
\usepackage{dcolumn}
\usepackage{bm}

\begin{document}

\preprint{APS/123-QED}

\title{Squeezing and Entanglement Dynamics in Phase-Sensitive Non-Hermitian Systems}

\author{Ruicong Huang}
 \altaffiliation[These authors]{ contributed equally to this work.}
\affiliation{%
Key Laboratory of Atomic and Subatomic Structure and Quantum Control (Ministry of Education), Guangdong Basic Research Center of Excellence for Structure and Fundamental Interactions of Matter, School of Physics, South China Normal University, Guangzhou 510006, China
}%
\affiliation{
Guangdong Provincial Key Laboratory of Quantum Engineering and Quantum Materials, Guangdong-Hong Kong Joint Laboratory of Quantum Matter, South China Normal University, Guangzhou 510006, China
}%
\author{Wencong Wang}%
 \altaffiliation[These authors]{ contributed equally to this work.}
 \affiliation{
 School of Electronic Science and Engineering (School of Microelectronics), South China Normal University, Foshan 528225, China 
}%
 \affiliation{
 Quantum Science Center of Guangdong-HongKong-Macao Greater Bay Area(Guangdong), Shenzhen, 518000, China
}%
\author{Yuyang Liang}
\affiliation{%
Key Laboratory of Atomic and Subatomic Structure and Quantum Control (Ministry of Education), Guangdong Basic Research Center of Excellence for Structure and Fundamental Interactions of Matter, School of Physics, South China Normal University, Guangzhou 510006, China
}%
\affiliation{
Guangdong Provincial Key Laboratory of Quantum Engineering and Quantum Materials, Guangdong-Hong Kong Joint Laboratory of Quantum Matter, South China Normal University, Guangzhou 510006, China
}%
\affiliation{
 Quantum Science Center of Guangdong-HongKong-Macao Greater Bay Area(Guangdong), Shenzhen, 518000, China
}%
\author{Dongmei Liu}
 \email{dmliu@scnu.edu.cn}
 \affiliation{
 School of Electronic Science and Engineering (School of Microelectronics), South China Normal University, Foshan 528225, China 
}%
 \affiliation{
 Quantum Science Center of Guangdong-HongKong-Macao Greater Bay Area(Guangdong), Shenzhen, 518000, China
}%
\author{Min Gu}
 \email{mingu@m.scnu.edu.cn}
 \affiliation{%
Key Laboratory of Atomic and Subatomic Structure and Quantum Control (Ministry of Education), Guangdong Basic Research Center of Excellence for Structure and Fundamental Interactions of Matter, School of Physics, South China Normal University, Guangzhou 510006, China
}%
\affiliation{
Guangdong Provincial Key Laboratory of Quantum Engineering and Quantum Materials, Guangdong-Hong Kong Joint Laboratory of Quantum Matter, South China Normal University, Guangzhou 510006, China
}%
\affiliation{
 Quantum Science Center of Guangdong-HongKong-Macao Greater Bay Area(Guangdong), Shenzhen, 518000, China
}%


\date{\today}

\begin{abstract}
Over the past decade, parity-time (PT) symmetry and anti-PT (APT) symmetry in various physical systems have been extensively studied, leading to significant experimental and theoretical advancements. However, physical systems that simultaneously exhibit both PT and APT symmetry have not yet been explored. In this study, we construct a phase-sensitive non-Hermitian wave mixing model that inherently possesses single-mode APT symmetry. By tuning the phase of the pump field, this model simultaneously exhibits two-mode quadrature-PT symmetry. Moreover, the APT phase transition is accompanied by the emergence of quantum entanglement from absence to presence. This remarkable quantum effect, quantum entanglement, which distinguishes itself from classical physics, is surprisingly linked to APT symmetry phase transition. Our work further explores the relationship between two-mode quantum entanglement and the phase of the pump field, offering deeper insight into the generation and evolution of entanglement in the corresponding nonlinear system. This provides a new perspective on quantum information processing. 
\end{abstract}

\maketitle



\emph{{Introduction}.---}While Hermitian operators have real eigenvalues, hermiticity is not required for real eigenvalues. Recent studies indicate that  non-Hermitian Hamiltonians with parity-time (PT) symmetry ($[H,PT]=0$) can possess real eigenvalues \cite{benderRealSpectraNonHermitian1998,benderMakingSenseNonHermitian2007,el-ganainyNonHermitianPhysicsPT2018}. These systems experience a phase transition where PT symmetry is spontaneously broken. At a specific parameter point, known as the exceptional point (EP), Hamiltonian eigenvalues shift from real to complex, and eigenvalues and eigenvectors coalesce, leading to novel phenomena \cite{ozdemirParityTimeSymmetry2019,bergholtzExceptionalTopologyNonHermitian2021,dingNonHermitianTopologyExceptionalpoint2022,wangNonHermitianOpticsPhotonics2023,chenExceptionalPointsEnhance2017a,wangObservationPTsymmetricQuantum2021,hanExceptionalEntanglementPhenomena2023}. Similarly, anti-PT (APT) symmetry, characterized by $\{H,PT\}=0$, also undergoes spontaneous symmetry breaking during the APT phase transition, with EPs emerging \cite{geAntisymmetricPTphotonicStructures2013,niuFourwaveMixingAntiparitytime2024}.\par
Over the past decade, PT symmetry \cite{el-ganainyTheoryCoupledOptical2007,guoObservationPTSymmetryBreaking2009,ruterObservationParityTime2010,schindlerExperimentalStudyActive2011,hangPTSymmetrySystem2013,pengParityTimesymmetricWhisperinggallery2014,wiersigEnhancingSensitivityFrequency2014,zhuPTSymmetricAcoustics2014,fengSinglemodeLaserParitytime2014,hodaeiParitytimeSymmetricMicroring2014,fleuryInvisibleAcousticSensor2015,liuMetrologyPTSymmetricCavities2016,zhangObservationParityTimeSymmetry2016} and APT symmetry \cite{pengAntiparityTimeSymmetry2016,liParitytimeSymmetryDiffusive2019,zhangDynamicallyEncirclingExceptional2019,jiangParityTimeSymmetricOptical2019,caoReservoirMediatedQuantumCorrelations2020,fanAntiparityTimeSymmetryPassive2020,bergmanObservationAntiparitytimesymmetryPhase2021,fengHarnessingDynamicalEncircling2022} have been extensively studied in fields like photonics, acoustics, and ultracold atoms, yielding notable experimental and theoretical progress. PT symmetry in physical systems typically requires linear gain and loss \cite{el-ganainyNonHermitianPhysicsPT2018,ozdemirParityTimeSymmetry2019,pengAntiparityTimeSymmetry2016,christodoulidesParitytimeSymmetryIts2018}. However, implementing PT symmetry in quantum systems is challenging due to Langevin noise \cite{scheelPTsymmetricPhotonicQuantum2018,zhangQuantumNoiseTheory2019,naghilooQuantumStateTomography2019}. Two approaches to address this are passive schemes based on loss modulation and Hamiltonian dilation \cite{naghilooQuantumStateTomography2019,wuObservationParitytimeSymmetry2019,klauckObservationPTsymmetricQuantum2019,yuExperimentalInvestigationQuantum2020,dingExperimentalDeterminationPTSymmetric2021,liuDynamicallyEncirclingExceptional2021}. Conversely, APT symmetry can be achieved without gain and loss \cite{jiangParityTimeSymmetricOptical2019,miriNonlinearityinducedPTsymmetryMaterial2016,wangNonHermitianDynamicsDissipation2019,luoQuantumSqueezingSensing2022}. Theoretical research suggests that APT symmetry in a two-mode bosonic system, free from linear gain and loss, can avoid Langevin noise and be implemented in nonlinear wave-mixing processes \cite{luoQuantumSqueezingSensing2022}. Recent studies on quantum entanglement \cite{einsteinCanQuantumMechanicalDescription1935,reidColloquiumEinsteinPodolskyRosenParadox2009} have shown unique PT and APT symmetry properties in two-qubit systems \cite{zhangExceptionalEntanglementQuantum2024} and linear optical systems \cite{fangEntanglementDynamicsAntiPTsymmetric2022}. However, no physical systems simultaneously displaying both PT and APT symmetry have been examined.

In this letter, we constructs a phase-sensitive non-Hermitian wave mixing model with inherent single-mode APT symmetry and achieves two-mode quadrature-PT symmetry by modulating the pump field phase, an aspect unexplored in prior research. Classical-nonclassical polarity (CNP) demonstrates the unique impact of APT symmetry on a system's nonclassical properties, significantly altering single-mode squeezing and two-body Gaussian entanglement dynamics as the system traverses EP. The inseparability criterion results provide insights into the relationship between two-mode quadrature-entanglement and the pump field phase, revealing the influence of single-mode APT and two-mode quadrature-PT symmetries on entanglement. Specifically, when $\cos\phi=0$, the system exhibits both single-mode APT and two-mode quadrature-PT symmetries, with two-mode quadrature-entanglement showing sudden death in the two-mode quadrature-PT-broken region and EP, and periodic revival in the two-mode quadrature-PT-symmetric region. Conversely, when $\cos\phi=\pm1$, the phase transition from APT-symmetric to APT-broken regions corresponds with the emergence of quantum entanglement from nonexistence, indicating a classical to quantum transition. This work further investigates the unique non-Hermitian properties of nonlinear systems, exploring the relationship between two-mode quantum entanglement and the pump field phase, elucidating the entanglement generation and evolution process in the model's nonlinear system, offering new perspectives for quantum information processing.

\begin{figure}
    \centering
    \includegraphics[width=8.5cm]{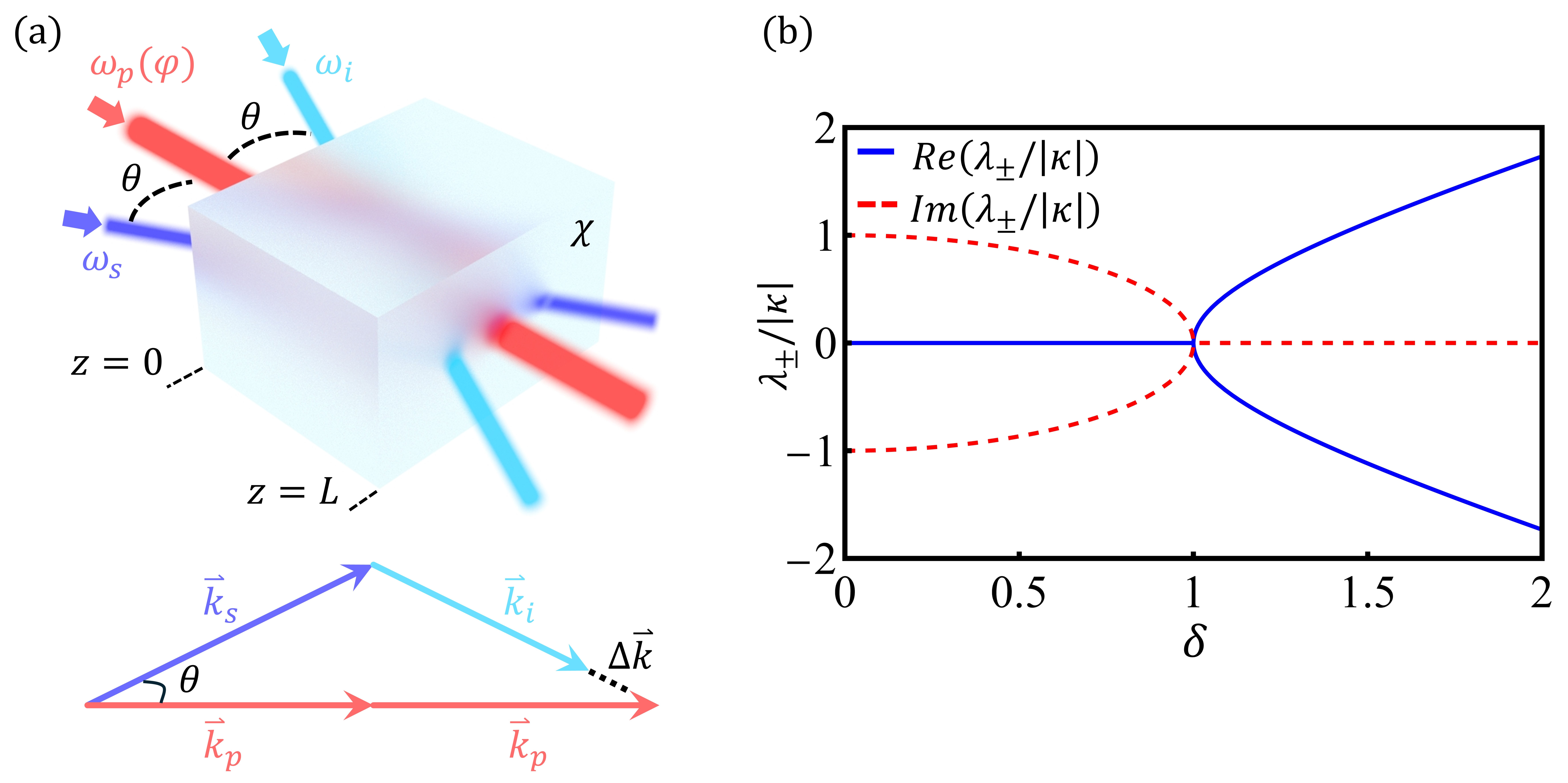}
    \caption{(a) A non-Hermitian system includes a four-wave mixing process and a phase-controllable pump source. Here, $\chi$ denotes the nonlinear susceptibility, $\varphi$ is the pump field phase, $\theta$ is the angle between the pump and signal (idler) fields, and $L$ is the total length of the nonlinear medium along the pump field's optical axis. 
    (b) APT phase transition process of the eigenvalues $\lambda_\pm$.}
    \label{fig:1}
\end{figure}

\emph{{Theoretical model}.---}Consider the Four-wave mixing process shown in Fig.~\ref{fig:1}(a). The input pump field $\omega_p$ with phase $\varphi$ and coherent state input signal field $\omega_s$ (idler field $\omega_i$) are incident onto a nonlinear medium (where $\chi$ represents the nonlinear susceptibility of the medium). During this process, two pump photons are annihilated and one signal photon and one idler photon are simultaneously generated. Considering an undepleted pump field, in the classical parametric approximation, the coupling equations between the signal field operator $a_s$ and idler field operator $a_i$ can be written as (we set $\hbar=1$)\cite{niuFourwaveMixingAntiparitytime2024,jiangParityTimeSymmetricOptical2019}
\begin{equation}
i\partial_z(a_s, a_i^{\dagger})^T=H_{\mathrm{single-mode}}(a_s,a_i^{\dagger})^T.
\label{eq:single-mode}
\end{equation}
 The non-Hermitian Hamiltonian matrix $H_{\mathrm{single-mode}}$ in Eq.~(\ref{eq:single-mode}) is
\begin{equation}
H_{\mathrm{single-mode}}=\left(\begin{array}{cc}
\Delta & \kappa e^{-i \phi} \\
-\kappa e^{i \phi} & -\Delta
\end{array}\right)
\label{eq:H_single-mode}
\end{equation}
where $\phi=2\varphi$, $\kappa$ is the real nonlinear coupling coefficient, and $\mathrm{\Delta}=-\Delta k/2$. As shown in Fig.~\ref{fig:1}(a), the phase mismatch $\Delta k=2k_p-\left(k_s+k_i\right)\cos\theta$, where $k_p$, $k_s$, and $k_i$ represent the wave numbers of the pump $(\omega_p)$, signal $(\omega_s)$, and idler fields $(\omega_i)$, respectively. Obviously, no matter the value of the phase $\phi$, $H_{single-mode}$ satisfies $\left\{H_{single-mode},PT\right\}=0$, where $P$ is the parity operator and $T$ is the time-reversal operator. This indicates that the single-mode system exhibits APT symmetry. Since there is no involvement of loss in $H_{single-mode}$, the Langevin noise operator does not need to be considered here to preserve the commutation relations $[a_j(0), a_j^{\dagger}(0)]=[a_j(L), a_j^{\dagger}(L)]=1(j=s, i)$. The eigenvalues of $H_{single-mode}$ are given by $\lambda_\pm=\pm\lambda=\pm i\left|\kappa\right|\sqrt{1-\delta^2}$, where $\delta=|\mathrm{\Delta}/\kappa|$. As shown in Fig.~\ref{fig:1}(b), when $\delta<1$, the eigenvalues are a pair of purely imaginary complex conjugates and the system is in the APT-symmetric region. When $\delta>1$, the eigenvalues are real and the system is in the APT-broken region. When $\delta=1$, the system is at the EP, where $\lambda_0=0$, indicating spontaneous symmetry breaking in the system.
Interestingly, by combining the quadrature of the optical field $q_j=\frac{1}{2}(a^{\dagger}_j+a_j)$ and $p_j=\frac{i}{2}(a^{\dagger}_j-a_j)$, where $\left[q_j,p_j\right]=\frac{i}{2}$, the dynamical equation between the two-mode quadrature can be rewritten as
\begin{subequations}
\begin{gather}
i \partial_z(q_s+q_i, p_s+p_i)^T=H_{\mathrm{two-mode(+)}}(q_s+q_i, p_s+p_i)^T,\label{eq:two-mode(+)}\\
i \partial_z(p_s-p_i, q_s-q_i)^T=H_{\mathrm{two-mode(-)}}(p_s-p_i, q_s-q_i)^T. \label{eq:two-mode(-)}
\end{gather}
\end{subequations}
In here, if and only if $\cos\phi=0$, the corresponding effective Hamiltonians in Eqs.~(\ref{eq:two-mode(+)}) and (\ref{eq:two-mode(-)}) are given by
\begin{subequations}
\begin{gather}
H_{two-mode(+)}=\left(\begin{array}{cc}
-i \kappa \sin \phi & i \Delta \\
-i \Delta & i \kappa \sin \phi
\end{array}\right) \label{Za}
\\
H_{two- mode(-)}=\left(\begin{array}{cc}
-i \kappa \sin \phi & -i \Delta \\
i \Delta & i \kappa \sin \phi
\end{array}\right) \label{Zb}
\end{gather}
\end{subequations}
which satisfying $\left[H_{two-mode(\pm)},PT\right]=0$, the two-mode effective Hamiltonian exhibits PT symmetry. The corresponding eigenvalues are given by $\lambda_{two-mode(\pm)}=\lambda_\pm=\pm\left|\kappa\right|\sqrt{\delta^2-1}$. When $\delta>1$, the eigenvalues are real and the system is in the two-mode quadrature-PT-symmetric region. When $\delta<1$, the eigenvalues are a pair of purely imaginary complex conjugates and the system is in the two-mode quadrature-PT-broken region. When $\delta=1$, the system is at the EP, where $\lambda_0=0$, indicating spontaneous symmetry breaking. The solutions to Eqs.~(\ref{eq:single-mode}) and (\ref{eq:two-mode(+)})-(\ref{eq:two-mode(-)}) can be found in the supplementary materials.

It is worth emphasizing that by satisfying a specific phase condition ($\cos\phi=0$), this phase-sensitive non-Hermitian wave mixing model can simultaneously exhibit single-mode APT symmetry and two-mode quadrature-PT symmetry, which has not been explored in previous research.

\emph{{Classical-Nonclassical Polarity}.---}To investigate the intrinsic connection between non-classical properties and the non-Hermitian nature of this phase-sensitive model, we first study Classical-Nonclassical
Polarity (CNP) \cite{liuClassicalNonclassicalPolarityGaussian2024} in our model. Using the covariance matrix $\gamma_{si}$ (see supplementary materials for more detail), the single-mode and two-mode CNPs can be written as \cite{liuClassicalNonclassicalPolarityGaussian2024}
\begin{subequations}
\begin{gather}
P_1=-\frac{\kappa^4 \sin ^4 \lambda L}{\lambda^4},\label{P1}\\
P_2=-\frac{\left(\kappa^2-2 \Delta^2+\kappa^2 \cos 2 \lambda L\right) \kappa^2 \sin ^2 \lambda L}{\lambda^4}. \label{P2}
\end{gather}
\end{subequations}
$P_1$ indicates the single-mode squeezing property of the system, with the signal and idler fields sharing identical single-mode squeezing characteristics. $P_2$ represents the two-mode Gaussian entanglement. CNP greater than 0 signifies a nonclassical state, CNP of 0 marks the boundary between classical and nonclassical states, and CNP less than 0 denotes a classical state. The absolute value of the CNP measures the distance from the classical-nonclassical boundary.

In here, the nonclassical properties measured by CNP exhibit two key characteristics. First, as indicated in Eq.~(\ref{P1}) and \eqref{P2}, the input coherent state's intensity does not affect the nonclassical properties measured by CNP. Second, the pump field's phase does not influence these properties, implying that single-mode squeezing and two-mode Gaussian entanglement remain unaffected by the pump field phase. The nonclassical properties mainly depend on the system parameters $\kappa$ and $\mathrm{\Delta}$, and the evolution distance $\kappa L$. Thus, discussing the CNP connection between single-mode APT symmetry and two-mode quadrature-PT symmetry here does not violate the condition $\cos \phi=0$.
\begin{figure}
    \centering
    \includegraphics[width=8.5cm]{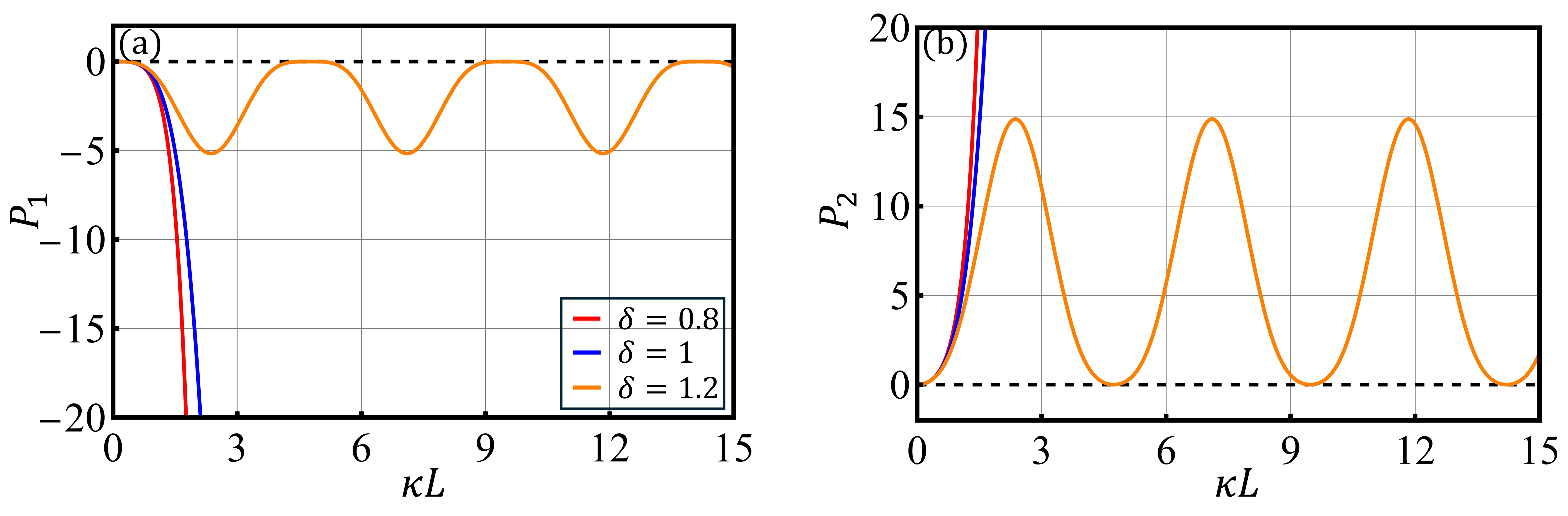}
    \caption{CNP of the system. The evolution of $P_1$ (a) and $P_2$ (b). The red solid line (for $\delta=0.8$) is in the APT-symmetric region, the blue solid line (for $\delta=1$) is at the EP, and the orange solid line (for $\delta=1.2$)) is in the APT-broken region.}
    \label{fig:2}
\end{figure}

Fig.~\ref{fig:2} (a) and (b) describe the evolution of $P_1$ and $P_2$, respectively. In fig.~\ref{fig:2}(a), regardless of the single-mode APT symmetry state, $P_1$ is less than or equal to $0$, indicating that the system does not exhibit single-mode squeezing, and thus demonstrates classical characteristics. Correspondingly, regardless of how the two-mode quadrature-PT symmetry varies, $P_2$ is greater than or equal to $0$ in fig.~\ref{fig:2}(b), indicating that the system consistently generates two-mode Gaussian entanglement throughout and reflecting nonclassical characteristics.More deeply, in single-mode APT-symmetric region and two-mode quadrature-PT-broken region ($\delta=0.8$), one of the eigenmodes disappears after a long evolution length owing to the presence of purely imaginary eigenvalues $\lambda_\pm$. The two-mode Gaussian entanglement measured by $P_2$ shows a monotonic increase. In the single-mode APT-broken region and two-mode quadrature-PT-symmetric region($\delta=1.2$), the two-mode Gaussian entanglement exhibits oscillatory behavior with a period of $T=\kappa\pi/\lambda$. At the EP ($\delta=1$), the two eigenmodes merge into a single mode. The two-mode Gaussian entanglement increases monotonically, but the rate of enhancement is slower compared to the single-mode APT-symmetric region and two-mode quadrature-PT-broken region. CNP reveals the unique connection between single-mode APT symmetry and two-mode quadrature-PT symmetry in the system's non-classical properties.

\emph{{The inseparability criterion}.---}To further explore the system's nonclassical properties, we consider another characterization of quantum entanglement: the inseparability criterion \cite{duanInseparabilityCriterionContinuous2000,simonPeresHorodeckiSeparabilityCriterion2000}. It reveals the relationship between system entanglement and the phase of the pump field and elucidates the impact of single-mode APT symmetry and two-mode quadrature-PT symmetry on entanglement.

The inseparability criterion is defined as follows: $E_1=Var\left(q_s-q_i\right)+Var\left(p_s+p_i\right)$, $E_2=Var\left(q_s+q_i\right)+Var\left(p_s-p_i\right)$. $E_1$ and $E_2$ are referred to as type-I and type-II entanglements of the system, respectively. $E_1\ <1$ or $E_2<1$ corresponds to the weaker criterion (wc). When $E_1\ <0.5$ or $E_2<0.5$, it corresponds to the stronger criterion (sc). The closer the values of $E_1$ and $E_2$ are to 0, the higher the degree of entanglement.

The theoretical calculation results are as follows:
\begin{subequations}
\begin{gather}
E_1=\cos ^2 \lambda L+\frac{\left(\kappa^2+\Delta^2+G1\right) \sin ^2 \lambda L+G2}{\lambda^2} \label{e1}
\\
E_2=\cos ^2 \lambda L+\frac{\left(\kappa^2+\Delta^2-G1\right) \sin ^2 \lambda L-G2}{\lambda^2} \label{e2}
\end{gather}
\end{subequations}
Where $G1=2 \kappa \Delta \cos \phi$ and $G2=\kappa \lambda \sin 2 \lambda z \sin \phi$. Consider the input field as coherent state. From Eqs. \eqref{e1} and \eqref{e2}, it is evident that the degree of entanglement is independent of the intensity of the input field. Instead, it is determined by the system parameters $\kappa$ and $\mathrm{\Delta}$ and the evolution distance $\kappa L$, and is also influenced by the phase of the pump field. Under different phase conditions (e.g., $\cos\phi=\pm1$ or $\cos\phi=0$), the system exhibits distinct entanglement evolution characteristics.

\begin{figure}
    \centering
    \includegraphics[width=8.5cm]{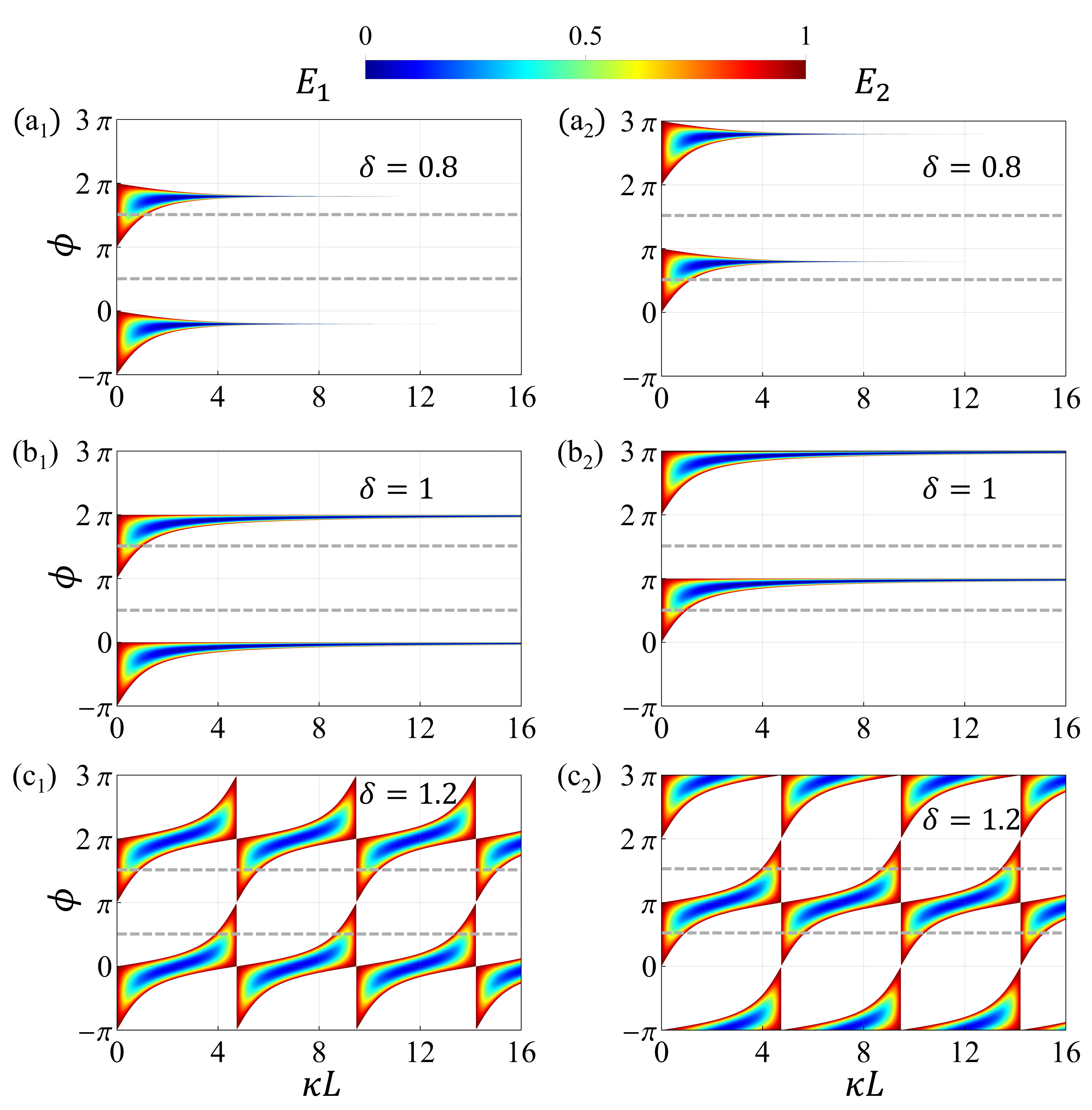}
    \caption{The evolution of the type-I/type-II entanglement under different phase $\phi$: ($a_1$), ($b_1$), and ($c_1$) correspond to type-I entanglement $E_1$; ($a_2$), ($b_2$), and ($c_2$) correspond to type-II entanglement $E_2$. ($a_1$) and ($a_2$) show the system in the APT-symmetric region with $\delta=0.8$; ($b_1$) and ($b_2$) show the system at the EP with $\delta=1$; ($c_1$) and ($c_2$) show the system in the APT-broken region with $\delta=1.2$. The dashed gray line indicates that the evolution simultaneously satisfies two-mode quadrature-PT symmetry($\cos \phi=0$).}
    \label{fig:3}
\end{figure}
Fig.\ref{fig:3} illustrates the evolution of type-I/type-II entanglement across different phases $\phi$. When $\cos\phi=0$ (e.g., $\phi=1.5\pi$ or $0.5\pi$), the system exhibits both single-mode APT symmetry and two-mode quadrature-PT symmetry, as indicated by the dashed gray line. Comparing fig.\ref{fig:3} ($\mathrm{a}_1$), ($b_1$), and ($c_1$) with ($a_2$), ($b_2$), and ($c_2$), reveals that type-I entanglement $E_1$ and type-II entanglement $E_2$ differ only by a phase of $\pi$. Thus, the analysis focuses on type-I entanglement $E_1$. Fig.\ref{fig:3}($a_1$) demonstrates that in the single-mode APT-symmetric region ($\delta=0.8$) and when $\pi<\phi<2\pi$, the system generates entanglement, which is maintained within a narrow phase range. Outside this range, entanglement appears briefly before quickly disappearing. For $0<\phi<\pi$, no entanglement is generated. Fig.\ref{fig:3}($c_1$) shows that in the single-mode APT-broken region ($\delta=1.2$), system entanglement evolves oscillatory with a period $T=\kappa\pi/\lambda$, and the entanglement degree can be adjusted by varying $\phi$. Notably, Fig.\ref{fig:3}($b_1$) indicates that at the EP ($\delta=1$), tuning $\phi$ close to 0 maintains entanglement over a broader phase range than in the APT-symmetric region. This highlights the significant role of EP control in non-Hermitian systems and the importance of $\phi$ tuning in generating entanglement in this model. \par

Specifically, when the system is at the EP ($\delta=1$), the expressions for $E_1$ and $E_2$ are simplified to
\begin{subequations}
\begin{gather}
E_1=1+2 \kappa^2 L^2-2 \kappa^2 L^2 \cos \phi+2 \kappa L \sin \phi \label{e11}
\\
E_2=1+2 \kappa^2 L^2+2 \kappa^2 L^2 \cos \phi-2 \kappa L \sin \phi \label{e22}
\end{gather}
\end{subequations}
It can be observed that when the phase condition $\cos\phi=1$ (or $\cos\phi=-1$) is met, $E_1=1$ (or $E_2=1$) is independent of the propagation distance, indicating that the system is in a critical state between the classical and quantum regimes.

\begin{figure}
    \centering
    \includegraphics[width=8.5cm]{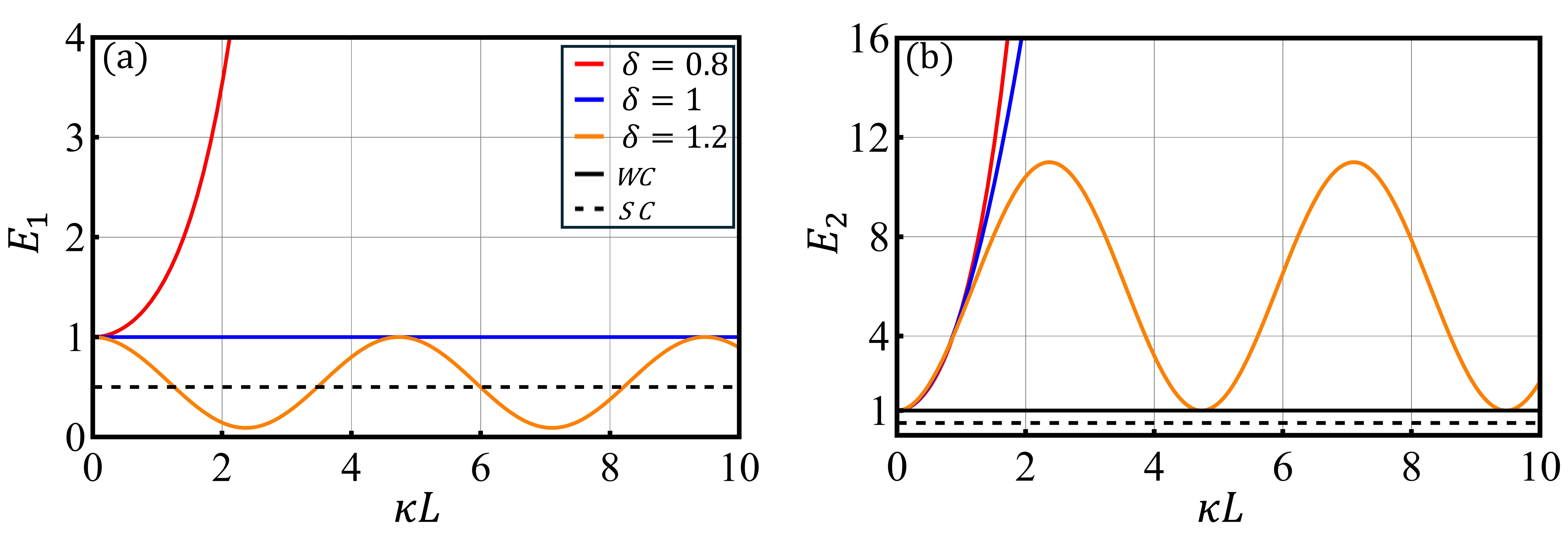}
    \caption{The evolution of entanglement when the pump phase satisfies $\cos\phi=1$. The solid red line represents the single-mode APT-symmetric region ($\delta=0.8$), the solid blue line indicates the EP ($\delta=1$), the solid orange line denotes the single-mode APT-broken region ($\delta=1.2$), the solid black line represents the weaker criterion (wc) for entanglement, and the dashed black line corresponds to the stronger criterion (sc) for entanglement. (a) Evolution of type-I entanglement. (b) Evolution of type-II entanglement.}
    \label{fig:4}
\end{figure}

Fig.~\ref{fig:4} clearly reveals the unique phenomenon that occurs in the system when the phase of the pump field satisfies the condition $\cos\phi=1$. As shown in Fig.~\ref{fig:4}(a), for type-I entanglement $E_1$, at the EP ($\delta=1$), $E_1=1$ represents a critical state between the classical and quantum regimes. No entanglement was observed in the single-mode APT-symmetric region ($\delta=0.8$). In the single-mode APT-broken region ($\delta=1.2$), the system consistently exhibits and maintains a certain degree of entanglement, with the entanglement degree oscillating with a period $T=\kappa\pi/\lambda$. The phase transition of the system from the APT-symmetric region through the EP to the APT-broken region is accompanied by a transition of quantum entanglement from nonexistence to presence, symbolizing a shift from classical to quantum behavior. The remarkable quantum effect of quantum entanglement, which distinguishes it from classical physics, is linked to APT symmetry phase transition. For type-II entanglement as shown in fig.~\ref{fig:4}(b), no entanglement is generated under this phase setting.

\emph{{Conclusion}.---}In summary, unlike previous studies, this study introduces a phase-sensitive non-Hermitian four-wave mixing model, wherein the single-mode effective Hamiltonian satisfies APT symmetry. Adjustment of the pump field phase enables the two-mode effective Hamiltonian to exhibit two-mode quadrature-PT symmetry. Analysis of the system's non-classical characteristics utilizing CNP and the inseparability criterion reveals that single-mode squeezing and two-mode Gaussian entanglement measured by CNP are phase-independent, while two-mode quadrature-entanglement measured by the inseparability criterion is phase-dependent. This distinction arises because CNP measures overall non-classicality, whereas the inseparability criterion evaluates quadrature component entanglement.

Theoretical calculations demonstrate that when $\cos\phi=0$, the system exhibits single-mode APT symmetry and two-mode quadrature-PT symmetry. Under these conditions, quantum entanglement experiences sudden death in the two-mode quadrature-PT-broken region and EP but exhibits periodic revival in the PT-symmetric region. When $\cos\phi=\pm1$, the system transitions from APT-symmetric through EP to APT-broken regions, with quantum entanglement emerging, signifying a shift from classical to quantum behavior. This research explores the unique non-Hermitian properties of nonlinear systems, elucidates the relationship between two-mode quantum entanglement and the pump phase, and enhances understanding of entanglement generation and evolution, offering novel insights for quantum information processing.

\emph{{Acknowledgments}.---}This work was supported by the Innovation Program for Quantum Science and Technology (Grant No.2021ZD0303401), the Quantum Science Strategic Project of Guangdong Province (Grant No.GDZX2306004), the National Natural Science Foundation of China under (No.Grant 62375089 and No.Grant 62471188), and Guangdong Basic and Applied Basic Research Foundation (Grant No.2022A1515140139).
\bibliography{ref}

\end{document}


\preprint{APS/123-QED}

\title{Supplementary Materials for "Squeezing and Entanglement Dynamics in Phase-Sensitive Non-Hermitian Systems"}

\author{Ruicong Huang}
 \altaffiliation[These authors]{ contributed equally to this work.}
\affiliation{%
Key Laboratory of Atomic and Subatomic Structure and Quantum Control (Ministry of Education), Guangdong Basic Research Center of Excellence for Structure and Fundamental Interactions of Matter, School of Physics, South China Normal University, Guangzhou 510006, China
}%
\affiliation{
Guangdong Provincial Key Laboratory of Quantum Engineering and Quantum Materials, Guangdong-Hong Kong Joint Laboratory of Quantum Matter, South China Normal University, Guangzhou 510006, China
}%
\author{Wencong Wang}%
 \altaffiliation[These authors]{ contributed equally to this work.}
 \affiliation{
 School of Electronic Science and Engineering (School of Microelectronics), South China Normal University, Foshan 528225, China 
}%
 \affiliation{
 Quantum Science Center of Guangdong-HongKong-Macao Greater Bay Area(Guangdong), Shenzhen, 518000, China
}%
\author{Yuyang Liang}
\affiliation{%
Key Laboratory of Atomic and Subatomic Structure and Quantum Control (Ministry of Education), Guangdong Basic Research Center of Excellence for Structure and Fundamental Interactions of Matter, School of Physics, South China Normal University, Guangzhou 510006, China
}%
\affiliation{
Guangdong Provincial Key Laboratory of Quantum Engineering and Quantum Materials, Guangdong-Hong Kong Joint Laboratory of Quantum Matter, South China Normal University, Guangzhou 510006, China
}%
\affiliation{
 Quantum Science Center of Guangdong-HongKong-Macao Greater Bay Area(Guangdong), Shenzhen, 518000, China
}%
\author{Dongmei Liu}
 \email{dmliu@scnu.edu.cn}
 \affiliation{
 School of Electronic Science and Engineering (School of Microelectronics), South China Normal University, Foshan 528225, China 
}%
 \affiliation{
 Quantum Science Center of Guangdong-HongKong-Macao Greater Bay Area(Guangdong), Shenzhen, 518000, China
}%
\author{Min Gu}
 \email{mingu@m.scnu.edu.cn}
 \affiliation{%
Key Laboratory of Atomic and Subatomic Structure and Quantum Control (Ministry of Education), Guangdong Basic Research Center of Excellence for Structure and Fundamental Interactions of Matter, School of Physics, South China Normal University, Guangzhou 510006, China
}%
\affiliation{
Guangdong Provincial Key Laboratory of Quantum Engineering and Quantum Materials, Guangdong-Hong Kong Joint Laboratory of Quantum Matter, South China Normal University, Guangzhou 510006, China
}%
\affiliation{
 Quantum Science Center of Guangdong-HongKong-Macao Greater Bay Area(Guangdong), Shenzhen, 518000, China
}%


\maketitle


\section{S1.Supplementary material on dynamical equation}
The solution to dynamical equation (1) can be expressed in the following matrix form:
\begin{widetext}
\begin{equation}
\binom{a_s(L)}{a_i^{\dagger}(L)}=\left(\begin{array}{cc}
\cos \lambda L-i \frac{\Delta}{\lambda} \sin \lambda L & -i e^{-i \phi} \frac{\kappa}{\lambda} \sin \lambda L \\
i e^{i \phi} \frac{\kappa}{\lambda} \sin \lambda L & \cos \lambda L+i \frac{\Delta}{\lambda} \sin \lambda L
\end{array}\right)\binom{a_s(0)}{a_i^{\dagger}(0)} \label{su}
\end{equation}
\end{widetext}

Throughout the dynamic process, it is straightforward to obtain $\left[a_j\left(L\right),{a_j}^\dag\left(L\right)\right]=\left[a_j\left(0\right),{a_j}^\dag\left(0\right)\right]=1 (j=s,i)$. The commutation relations of the modes are consistently preserved.

By combining the quadrature of the optical field $q_j=\frac{1}{2}\left(a_j+{a_j}^\dag\right)$ and $p_j=-\frac{i}{2}\left(a_j-{a_j}^\dag\right)$, where $j=s,i$, and $\left[q_j,p_j\right]=\frac{i}{2}$. Equation (1) can be transformed into
\begin{widetext}
\begin{subequations}
\begin{gather}
i \partial_z\binom{q_s+q_i}{p_s+p_i}=i\left(\begin{array}{cc}
-\kappa \sin \phi & \Delta-\kappa \cos \phi \\
-\Delta-\kappa \cos \phi & \kappa \sin \phi
\end{array}\right)\binom{q_s+q_i}{p_s+p_i} \label{aa}
\\
i \partial_z\binom{p_s-p_i}{q_s-q_i}=i\left(\begin{array}{cc}
-\kappa \sin \phi & -\Delta+\kappa \cos \phi \\
\Delta+\kappa \cos \phi & \kappa \sin \phi
\end{array}\right)\binom{p_s-p_i}{q_s-q_i} \label{bb}
\end{gather}
\end{subequations}
\end{widetext}
The solutions to dynamical equations \eqref{aa} and \eqref{bb} can be expressed in the following matrix form:
\begin{widetext}
\begin{subequations}
\begin{gather}
\binom{q_s(L)+q_i(L)}{p_s(L)+p_i(L)}=\left(\begin{array}{cc}
\cos \lambda L-\frac{\kappa}{\lambda} \sin \lambda L \sin \phi & \frac{\Delta}{\lambda} \sin \lambda L-\frac{\kappa}{\lambda} \sin \lambda L \cos \phi \\
-\frac{\Delta}{\lambda} \sin \lambda L-\frac{\kappa}{\lambda} \sin \lambda L \cos \phi & \cos \lambda L+\frac{\kappa}{\lambda} \sin \lambda L \sin \phi
\end{array}\right)\binom{q_s(0)+q_i(0)}{p_s(0)+p_i(0)}
\\
\binom{p_s(L)-p_i(L)}{q_s(L)-q_i(L)}=\left(\begin{array}{cc}
\cos \lambda L-\frac{\kappa}{\lambda} \sin \lambda L \sin \phi & -\frac{\Delta}{\lambda} \sin \lambda L+\frac{\kappa}{\lambda} \sin \lambda L \cos \phi \\
\frac{\Delta}{\lambda} \sin \lambda L+\frac{\kappa}{\lambda} \sin \lambda L \cos \phi & \cos \lambda L+\frac{\kappa}{\lambda} \sin \lambda L \sin \phi
\end{array}\right)\binom{p_s(0)-p_i(0)}{q_s(0)-q_i(0)}
\end{gather}
\end{subequations}
\end{widetext}

\section{S2.Single-Mode Quadrature-Variance and Logarithmic Negativity}
To verify the single-mode CNP, we calculated the single-mode variances,  $Var\left(q_s\right)$, $Var\left(p_s\right)$, $Var\left(q_i\right)$, and $Var\left(p_i\right)$. The variance is defined as $Var\left(X\right)=\left\langle X^2\right\rangle-\left\langle X\right\rangle^2$.

It is straightforward to obtain
\begin{equation}
Var\left(p_s\right)=\ Var\left(q_i\right)=\ Var\left(p_i\right)=Var\left(q_s\right)=\frac{\mathrm{\Delta}^2-\kappa^2\cos{\left(\ 2L\sqrt{\mathrm{\Delta}^2-\kappa^2}\right)}}{4\left(\mathrm{\Delta}^2-\kappa^2\right)} \label{sq}
\end{equation}

Equation \eqref{sq} shows that the single-mode quadrature variances are independent of the input field intensity and the phase of the pump field.

\begin{figure}
    \centering
    \includegraphics[width=10cm]{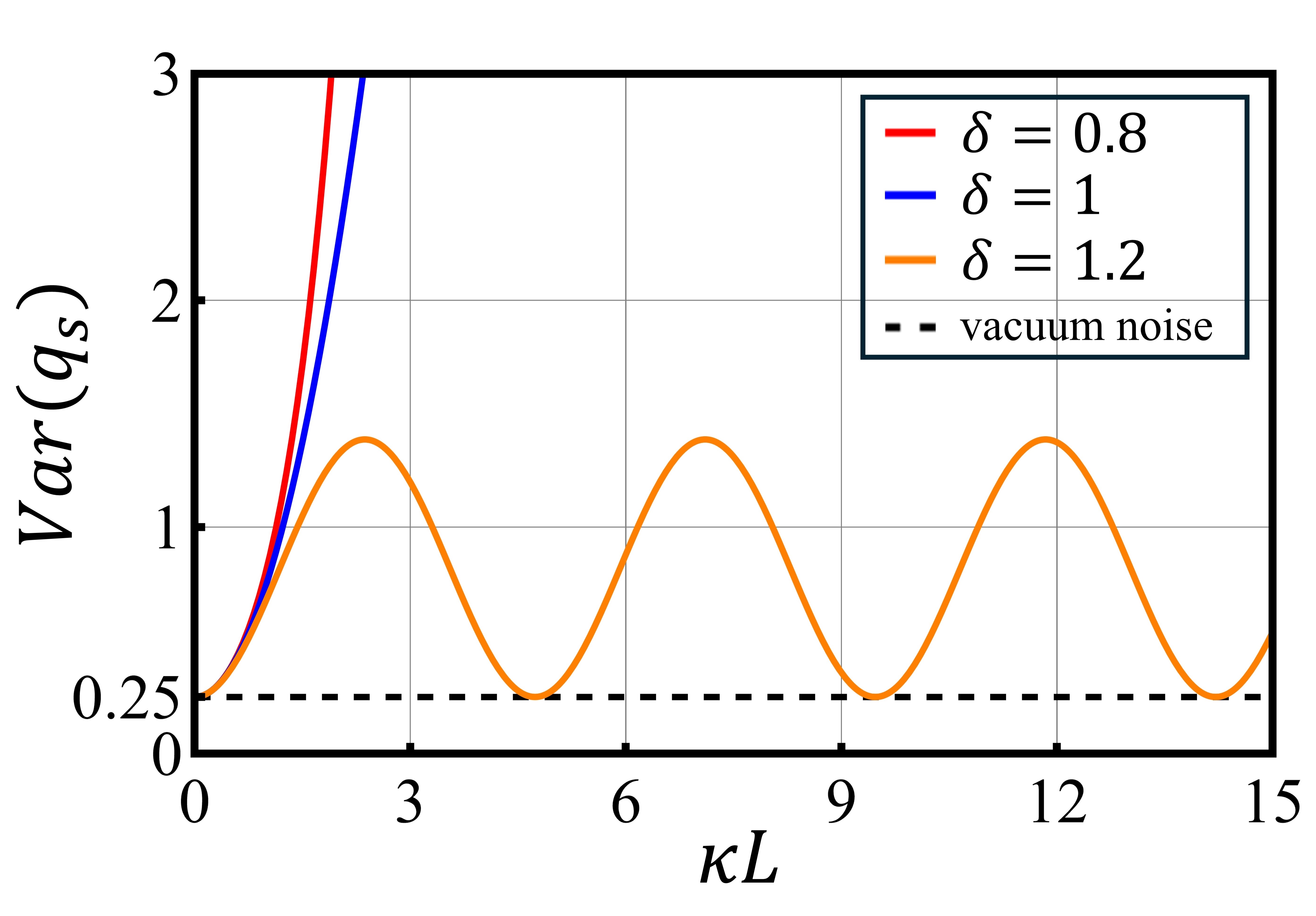}
    \caption{Evolution of single-mode quadrature-variance $Var\left(q_s\right)$ for different values of $\delta$. The solid red line ($\delta=0.8$) represents the APT-symmetric region, solid blue line ($\delta=1$) represents the EP, and solid orange line ($\delta=1.2$) represents the APT-broken region. The dashed black line represents the vacuum noise level.}
\end{figure}
Figure S1 shows the evolution of $Var\left(q_s\right)$. $Var\left(q_s\right)$ is greater than the vacuum noise in the APT-symmetric region, EP, and APT-broken region, indicating that the system does not exhibit single-mode squeezing and thus reflects classical characteristics. This corroborates the results of single-mode CNP discussed in the main text.

To verify the two-mode CNP, we calculate logarithmic negativity \cite{vidalComputableMeasureEntanglement2002,plenioLogarithmicNegativityFull2005} as an entanglement measure. This calculation requires the following covariance matrix of the system:
\begin{equation}
V=\left(\begin{matrix}A&C\\C^T&B\\\end{matrix}\right) \label{cv}
\end{equation}
Where,
\begin{equation}
V_{nm}=\frac{1}{2}\left\langle X_nX_m+X_mX_n\right\rangle-\left\langle X_n\right\rangle\left\langle X_m\right\rangle \label{v}
\end{equation}
Define $X\left(z\right)=\left[q_s\left(z\right),p_s\left(z\right),q_i\left(z\right),p_i\left(z\right)\right]^T$. The input signal and idler fields are initially in coherent states. The quantum states evolving according to the dynamical equations remain in Gaussian states, and all properties of these Gaussian states are encapsulated in the covariance matrix $V$. The logarithmic negativity can be calculated using the covariance matrix $V$ as follows: $E_N=\max{\left[\ 0,-\ln{4\eta}\right]}$, where $\eta=\frac{1}{\sqrt2}\sqrt{\mathrm{\Sigma}-\sqrt{\mathrm{\Sigma}^2-4\det{V}}}$  and $\mathrm{\Sigma}=\det{A}+\det{B}-2\det{C}$. $E_N>0$ indicates entanglement, with larger values of $E_N$ corresponding to stronger entanglement.

The entanglement measured by logarithmic negativity has two important properties. First, as indicated by Equation \eqref{v}, the first and second terms on the right-hand side of the equation depend on the field operators. The difference between these terms ensures that the matrix elements of covariance matrix $V$ are independent of the intensity of the input fields. Second, the entanglement measured by logarithmic negativity is also independent of the pump phase. The degree of entanglement primarily depends on the system parameters $\kappa$ and $\mathrm{\Delta}$, as well as the evolution distance $\kappa L$.

\begin{figure}
    \centering
    \includegraphics[width=16cm]{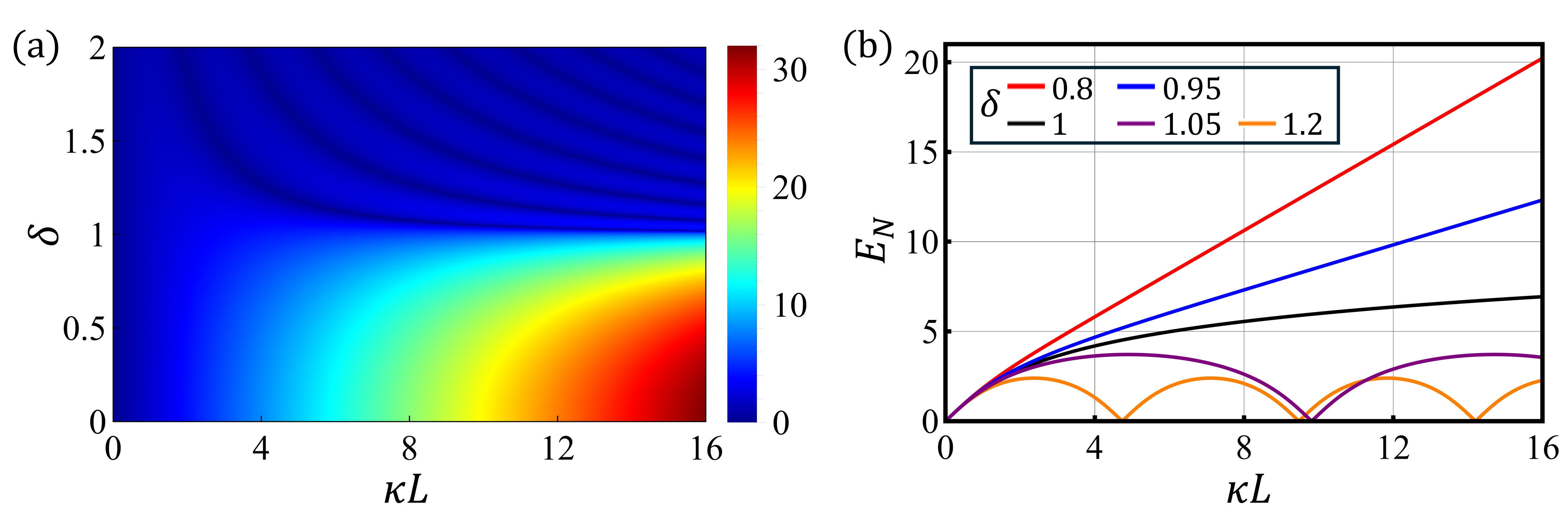}
    \caption{Logarithmic negativity for the system (a) Variation of $E_N$ with evolution distance $\kappa L$ and system parameter $\delta$. (b) Evolution of $E_N$ for different $\delta$ values The red solid line ($\delta=0.8$) and the blue solid line ($\delta=0.95$) correspond to the APT-symmetric region, the black solid line ($\delta=1$) corresponds to the EP, and the purple solid line ($\delta=1.05$) and the orange solid line ($\delta=1.2$) correspond to the APT-broken region.}
\end{figure}
Figure S2(a) shows the entanglement results (measured by the logarithmic negativity $E_N$) as a function of the system parameter $\delta$ and the evolution distance $\kappa L$. It is evident that significant changes in the entanglement behavior occur near the EP ($\delta=1$). To illustrate the entanglement evolution more clearly, Figure S2(b) shows the entanglement dynamics for $\delta=0.8$, $0.95$, $1$, $1.05$, and $1.2$. The results indicate that entanglement exhibits different evolution patterns in the APT-symmetric region, EP, and APT-broken region. In the APT-symmetric region ($\delta<1$), owing to the presence of pure imaginary eigenvalues $\lambda_\pm$, one of the eigenmodes disappears after prolonged evolution, resulting in a monotonic increase in the entanglement of the system. Moreover, as the value of $\delta$ decreases (where $\delta=\left|\mathrm{\Delta}/\kappa\right|$), assuming that $\kappa$ is fixed and $\mathrm{\Delta}$ decreases), the rate of entanglement enhancement in the system also increases. When $\delta\rightarrow0$ ($\mathrm{\Delta}\rightarrow0$), the rate of entanglement enhancement reaches its maximum, which corresponds to a perfect phase-matching condition in the nonlinear process. In the APT-broken region ($\delta>1$), the entanglement exhibits an oscillatory behavior with a period $T=\kappa\pi/\lambda$. The closer $\delta$ is to the EP, the longer the oscillation period and the stronger the achievable entanglement. At the EP ($\delta=1$), the two eigenmodes merge into a single mode, leading to a monotonic increase in entanglement, although the rate of enhancement is slower than that in the APT-symmetric region. In summary, the entanglement measured by logarithmic negativity corroborates the results obtained from the two-mode CNP discussed in the main text.

\section{S3.A Simple Verification of Commutation Relations}
Through algebraic manipulations, it can be verified that the solutions provided in Equation \eqref{su} consistently maintain the commutation relations
\begin{equation}
\left[a_s\left(L\right),{a_s}^\dag\left(L\right)\right]=\left(\cos^2{\lambda}L+\frac{\mathrm{\Delta}^2\sin^2{\lambda}L}{\lambda^2}\right)\left[a_s\left(0\right),{a_s}^\dag\left(0\right)\right]-\frac{\kappa^2\sin^2{\lambda}L}{\lambda^2}\left[a_i\left(0\right),{a_i}^\dag\left(0\right)\right]=1
\end{equation}

\section{S4.Calculation of CNP}
To calculate CNP\cite{liuClassicalNonclassicalPolarityGaussian2024}, a system covariance matrix is required. The covariance matrix of the system can be expressed as
\begin{equation}
\gamma_{s i}=\left(\begin{array}{cc}
\gamma_s & x \\
x^{\dagger} & \gamma_i
\end{array}\right)
\end{equation}
Where,
\begin{equation}
\gamma_{n m}=\frac{1}{2}\left\langle\Delta r_n \Delta r_m{ }^{\dagger}+\Delta r_m{ }^{\dagger} \Delta r_n\right\rangle
\end{equation}
Define $r=\left[{a_s}^\dag,a_s,{a_i}^\dag,a_i\right]^\dag$ and $\Delta r_n=r_n-\left\langle r\right\rangle_n$. The initial states of both the input signal field and idler field under study are in the coherent state. According to the dynamical equations, the evolving quantum state remains in a Gaussian state, and all properties of this Gaussian state are encapsulated in the covariance matrix $\gamma_{si}$.
\begin{equation}
\gamma_{si}=\left(\begin{matrix}\frac{\mathrm{\Delta}^2-\kappa^2\cos{2\lambda L}}{2\left(\mathrm{\Delta}^2-\kappa^2\right)}&0&0&-\frac{e^{-i\phi}\kappa\left(\mathrm{\Delta}+i\lambda\cot{\lambda L}\right)\sin^2{\lambda L}}{\mathrm{\Delta}^2-\kappa^2}\\0&\frac{\mathrm{\Delta}^2-\kappa^2\cos{2\lambda L}}{2\left(\mathrm{\Delta}^2-\kappa^2\right)}&\frac{e^{i\phi}\kappa\left(-\mathrm{\Delta}+i\lambda\cot{\lambda L}\right)\sin^2{\lambda L}}{\mathrm{\Delta}^2-\kappa^2}&0\\0&\frac{e^{-i\phi}\kappa\left(-\mathrm{\Delta}-i\lambda\cot{\lambda L}\right)\sin^2{\lambda L}}{\mathrm{\Delta}^2-\kappa^2}&\frac{\mathrm{\Delta}^2-\kappa^2\cos{2\lambda L}}{2\left(\mathrm{\Delta}^2-\kappa^2\right)}&0\\-\frac{e^{i\phi}\kappa\left(\mathrm{\Delta}-i\lambda\cot{\lambda L}\right)\sin^2{\lambda L}}{\mathrm{\Delta}^2-\kappa^2}&0&0&\frac{\mathrm{\Delta}^2-\kappa^2\cos{2\lambda L}}{2\left(\mathrm{\Delta}^2-\kappa^2\right)}\\\end{matrix}\right)
\end{equation}
Using the covariance matrix $\gamma_{si}$, the single-mode and two-mode CNPs can be written as
\begin{equation}
P_1=-\left(v-\frac{1}{2}\right)\left(N-\frac{1}{2}\right)=-\frac{\kappa^4 \sin ^4 \lambda L}{\lambda^4}
\end{equation}
\begin{equation}
P_2=-\frac{1}{8}+\frac{1}{2} I_1-2 I_0=-\frac{\left(\kappa^2-2 \Delta^2+\kappa^2 \cos 2 \lambda L\right) \kappa^2 \sin ^2 \lambda L}{\lambda^4}
\end{equation}
Let $\nu$ and $N$ be the eigenvalues of the matrices $\gamma_s$ and $\gamma_i$, respectively. The single- and two-mode CNPs are then given by $I_1=|\gamma_s|+|\gamma_i|-2|x|$, $I_0=|\gamma_{si}|$. 

\section{S5.Calculation of the Separability Criterion}
Through some straightforward algebraic manipulations, we obtain
\begin{subequations}
\begin{gather}
Var\left(q_s-q_i\right)=Var\left(p_s+p_i\right)=\frac{\left(\mathrm{\Delta}+\kappa\cos{\phi}\right)^2\sin^2{\lambda L}+\left(\lambda\cos{\lambda L}+\kappa\sin{\lambda L}\sin{\phi}\right)^2}{2\lambda^2}
\\
Var\left(q_s+q_i\right)=Var\left(p_s-p_i\right)=\frac{\left(\mathrm{\Delta}-\kappa\cos{\phi}\right)^2\sin^2{\lambda L}+\left(\lambda\cos{\lambda L}-\kappa\sin{\lambda L}\sin{\phi}\right)^2}{2\lambda^2} 
\end{gather}
\end{subequations}

By combining the definitions of the inseparability criteria, $E_1=Var\left(q_s-q_i\right)+Var\left(p_s+p_i\right)$ and $E_2=Var(q_s+q_i\ )+Var(p_s-p_i\ )$, the results for Equations (6a) and (6b) in the main text can be derived.
